\documentclass[12pt]{article}
\usepackage{enumerate}
\usepackage{amsfonts}
\usepackage{amsmath}
\usepackage{amssymb}
\usepackage{amsthm}

\def\H{\mathcal{H}}

\def\P{\mathcal{P}}
\def\S{\mathfrak{S}}
\def\T{\mathfrak{T}}
\def\B{\mathfrak{B}}

\newcommand{\id}{\mathrm{Id}}
\newcommand{\Tr}{\mathrm{Tr}}

\newcounter{defin}  \newcounter{lemma}  \newcounter{theorem}
\newcounter{property} \newcounter{corol}  \newcounter{remark} \newcounter{example}

\newenvironment{lemma}{\par\refstepcounter{lemma}
     \textbf{Lemma \thelemma.} }{\rm\par}
\newenvironment{theorem}{\par\refstepcounter{theorem}
     \textbf{Theorem \thetheorem.}\ }{\rm\par}
\newenvironment{property}{\par\refstepcounter{property}
     \textbf{Proposition \theproperty.}\ }{\rm\par}
\newenvironment{corollary}{\par\refstepcounter{corol}
     \textbf{Corollary \thecorol.} }{\rm\par}
\newenvironment{definition}{\par\refstepcounter{defin}
     \textbf{Definition \thedefin.}\ }{\rm\par}
\newenvironment{remark}{\par\refstepcounter{remark}
     \textbf{Remark \theremark.}}{\rm\par}

\begin{document}
\title{A criterion for coincidence of the entanglement-assisted classical capacity
and the Holevo capacity of a quantum channel.\footnote{This work was
partially supported by the
scientific program ``Mathematical control theory'' of RAS.}}
\author{M.E. Shirokov\\
Steklov Mathematical Institute, RAS, Moscow\\
msh@mi.ras.ru}
\date{}
\maketitle

\begin{abstract}

It is easy to show coincidence of the entanglement-assisted classical capacity
and the Holevo capacity for any c-q channel between finite dimensional quantum systems. In this paper we
prove the converse assertion:  coincidence of the above-mentioned
capacities of a quantum
channel implies that the $\chi$-essential part of this channel is a
c-q channel (the $\chi$-essential part is a restriction
of a channel obtained by discarding all input states useless for
transmission of classical information).


The above observations are generalized to infinite dimensional quantum channels with linear constraints by using the obtained conditions
for coincidence of the quantum mutual information and the constrained Holevo capacity.

\end{abstract}
\maketitle

\tableofcontents

\section{Introduction}

In this paper we specify the necessary and sufficient conditions for coincidence of the Holevo capacity $\bar{C}(\Phi)$
and the entanglement-assisted (classical) capacity
$C_{\mathrm{ea}}(\Phi)$ of a quantum channel $\Phi$ presented in \cite{Sh-16}
and generalize these conditions to the infinite dimensional case by using the results concerning reversibility of a quantum channel with respect to families of pure states obtained in \cite{Sh-18}.

In contrast to an intuitive point of view,  the class of channels $\Phi$ such
that
\begin{equation}\label{coin}
\bar{C}(\Phi)=C_{\mathrm{ea}}(\Phi)
\end{equation}
does not coincide with the class of entanglement-breaking channels.
The examples of entanglement-breaking channels, in particular, of
q-c channels, for which $\,\bar{C}(\Phi)<C_{\mathrm{ea}}(\Phi),\,$ can
be found in \cite{BSST,H-CQM,Sh-16}. In \cite{BSST+} the non-entanglement-breaking
channel such that equality (\ref{coin}) holds is constructed. A criterion of this equality for the class of q-c
channels defined by quantum observables is recently obtained in \cite{H-CQM}.

In the first part of the paper we show that equality (\ref{coin}) holds for a finite dimensional quantum channel $\Phi$
if (correspondingly, only if) this channel (correspondingly, the $\chi$-essential part of this channel) belongs to the class of c-q channels (the $\chi$-essential part is defined as a
restriction of a channel to the set of states supported by the
minimal subspace containing elements of \emph{all} ensembles optimal
for this channel in the sense of the Holevo capacity, see Definition
\ref{ef-subch}). Thus, assuming  that we deal with channels
reduced to their $\chi$-essential parts, we may say that the class
of channels, for which equality (\ref{coin}) holds, coincides with the class of c-q channels.

In the second part of the paper the above observations are generalized to infinite dimensional channels with linear constraints by studying the equality conditions for the general inequality
\begin{equation}\label{g-ineq}
\bar{C}(\Phi,\rho)\leq I(\Phi,\rho),
\end{equation}
connecting the constrained Holevo capacity $\bar{C}(\Phi,\rho)$ and the quantum mutual
information $I(\Phi,\rho)$ of a quantum channel $\Phi$ at a state $\rho$. It is shown that the equality in (\ref{g-ineq}) implies
that the restriction of the channel $\Phi$ to the set of states
supported by the subspace $\,\mathrm{supp}\rho\,$ (the support of $\rho$) is
a c-q channel. The converse implication holds if the state $\rho$ is diagonizable in the basis from the c-q representation of the above restriction of the channel $\Phi$.

\bigskip

\section{Preliminaries}

Let $\H$ be either a finite dimensional or
separable Hilbert space, $\B(\H)$ and
$\mathfrak{T}( \mathcal{H})$ -- the Banach spaces of all bounded
operators in $\mathcal{H}$ and of all trace-class operators in
$\H$  correspondingly, $\S(\H)$ -- the closed convex subset
of $\mathfrak{T}(\H)$ consisting of positive operators
with unit trace called \emph{states} \cite{N&Ch}.

Denote by $I_{\mathcal{H}}$ and $\mathrm{Id}_{\mathcal{H}}$ the
unit operator in a Hilbert space $\mathcal{H}$ and the identity
transformation of the Banach space $\mathfrak{T}(\mathcal{H})$
correspondingly.

A linear completely positive trace preserving map
$\Phi:\mathfrak{T}(\mathcal{H}_A)\rightarrow\mathfrak{T}(\mathcal{H}_B)$
is called  \emph{quantum channel} \cite{N&Ch}.
The Stinespring
theorem implies existence of a Hilbert space $\mathcal{H}_E$ and of
an isometry
$V:\mathcal{H}_A\rightarrow\mathcal{H}_B\otimes\mathcal{H}_E$ such
that
\begin{equation}\label{Stinespring-rep}
\Phi(A)=\mathrm{Tr}_{\mathcal{H}_E}VA V^{*},\quad
A\in\mathfrak{T}(\mathcal{H}_A).
\end{equation}
The quantum  channel
\begin{equation}\label{c-channel}
\mathfrak{T}(\mathcal{H}_A)\ni
A\mapsto\widehat{\Phi}(A)=\mathrm{Tr}_{\mathcal{H}_B}VAV^{*}\in\mathfrak{T}(\mathcal{H}_E)
\end{equation}
is called \emph{complementary} to the channel $\Phi$
\cite{H-comp-ch}.\footnote{The quantum channel $\widehat{\Phi}$ is
also called \emph{conjugate} to the channel $\Phi$ \cite{KMNR}.} The
complementary channel is defined uniquely (cf.\cite[the Appendix]{H-comp-ch}): if
$\widehat{\Phi}':\mathfrak{T}(\mathcal{H}_A)\rightarrow\mathfrak{T}(\mathcal{H}_{E'})$
is a channel defined by (\ref{c-channel}) via the Stinespring isometry
$V':\mathcal{H}_A\rightarrow\mathcal{H}_B\otimes\mathcal{H}_{E'}$ then
the channels $\widehat{\Phi}$ and $\widehat{\Phi}'$ are \emph{isometrically equivalent} in the sense that
there is a partial isometry
$W:\mathcal{H}_E\rightarrow\mathcal{H}_{E'}$ such that
\begin{equation}\label{c-isom}
\widehat{\Phi}'(A)=W\widehat{\Phi}(A)W^*,\quad\widehat{\Phi}(A)=W^*\widehat{\Phi}'(A)W,\quad
A\in \T(\H_A).
\end{equation}

A channel $\Phi:\mathfrak{T}(\mathcal{H}_A)\rightarrow\mathfrak{T}(\mathcal{H}_B)$ is called \emph{classical-quantum} (briefly a \emph{c-q channel}) if it has the following representation
\begin{equation}\label{c-q-rep}
\Phi(\rho)=\sum_{k=1}^{\dim\H_A}\langle k|\rho|k\rangle\sigma_k,
\end{equation}
where $\{|k\rangle\}$ is an orthonormal basis in $\H_A$ and
$\{\sigma_k\}$ is a collection of states in $\S(\H_B)$ \cite{N&Ch}.
\smallskip

Let $H(\rho)$ and $H(\rho\|\sigma)$ be respectively the von Neumann
entropy of the state $\rho$ and the quantum relative entropy of the
states $\rho$ and $\sigma$ \cite{L,N&Ch}. Let
$$
\chi(\{\pi_i,\rho_i\})\doteq\sum_i\pi_i
H(\rho_i\|\bar{\rho})=H(\bar{\rho})-\sum_i\pi_i
H(\rho_i)
$$
be the $\chi$-quantity of an ensemble $\{\pi_i,\rho_i\}$ of quantum states with the average state $\bar{\rho}$, where the second expression is valid under the condition $H(\bar{\rho})<+\infty$.

\medskip

The constrained Holevo capacity of the channel $\Phi$ at a state $\rho\in\S(\H_A)$ is defined as
follows
\begin{equation}\label{chi-fun}
\bar{C}(\Phi,\rho)\doteq\sup_{\{\pi_i,\rho_i\},\,\bar{\rho}=\rho}\chi(\{\pi_i,\Phi(\rho_i)\})
\end{equation}
(the supremum is over all ensembles of states in $\S(\H_A)$ with the average state $\rho$).
If $H(\Phi(\rho))<+\infty$ then
\begin{equation}\label{chi-fun+}
\bar{C}(\Phi,\rho)=H(\Phi(\rho))-\hat{H}_{\Phi}(\rho),
\end{equation}
where $\hat{H}_{\Phi}(\rho)=\inf_{\{\pi_i,\rho_i\},\,\bar{\rho}=\rho}\sum_i\pi_i
H(\Phi(\rho_i))$ (the infimum here can be taken over ensembles of pure states by concavity of the function $\rho\mapsto
H(\Phi(\rho))$).\medskip

In finite dimensions the quantum mutual information of the channel $\Phi$ at a state $\rho\in\S(\H_A)$ is defined as
follows (cf.\cite{N&Ch})
\begin{equation}\label{mi-rep+}
I(\Phi,\rho)=H(\rho)+H(\Phi(\rho))-H(\widehat{\Phi}(\rho)).
\end{equation}
Since in infinite dimensions the terms in the right side of
(\ref{mi-rep+}) may be infinite, it is reasonable to define the
quantum mutual information by the following formula
$$
I(\Phi,\rho) = H\left(\Phi \otimes \id_{R}
(|\varphi_{\rho}\rangle\langle\varphi_{\rho}|)\, \|\, \Phi (\rho)
\otimes \varrho\right),
$$
where $\varphi_{\rho}$ is a purification vector\footnote{This means
that $\Tr_{\H_R}|\varphi_{\rho}\rangle\langle\varphi_{\rho}|=\rho$.}
in $\H_A\otimes\H_R$ for the state $\rho\in\S(\H_A)$ and
$\varrho=\Tr_{\H_A}|\varphi_{\rho}\rangle\langle\varphi_{\rho}|$ is
a state in $\S(\H_R)$ isomorphic to $\rho$. If $H(\rho)$ and
$H(\Phi(\rho))$ are finite then the last formula for $I(\Phi,\rho)$
coincides with (\ref{mi-rep+}).\medskip

If $\rho$ is a state with finite entropy then the quantum mutual information of an arbitrary channel $\Phi$ at the state $\rho$ can be expressed as follows
\begin{equation}\label{mi-rep++}
I(\Phi,\rho)=H(\rho)+\bar{C}(\Phi,\rho)-\bar{C}(\widehat{\Phi},\rho)=\bar{C}(\Phi,\rho)+\Delta_{\Phi}(\rho),
\end{equation}
where
$\Delta_{\Phi}(\rho)=H(\rho)-\bar{C}(\widehat{\Phi},\rho)\geq0$ (by
monotonicity of the relative entropy). If $H(\Phi(\rho))$ and $H(\widehat{\Phi}(\rho))$ are finite then expression (\ref{mi-rep++}) can be easily derived from (\ref{chi-fun+}) and (\ref{mi-rep+}), since
$\hat{H}_{\Phi}\equiv\hat{H}_{\widehat{\Phi}}$ (this follows from
coincidence of the functions $\rho\mapsto H(\Phi(\rho))$ and
$\rho\mapsto H(\widehat{\Phi}(\rho))$ on the set of pure states), in general case it can be proved by using Lemma
\ref{inf-dim-l} in the Appendix.
\medskip

\section{Finite dimensional channels}

The Holevo capacity of a finite dimensional channel $\Phi:\S(\H_A)\rightarrow\S(\H_B)$ is defined as follows
\begin{equation}\label{chi-cap}
\bar{C}(\Phi)\doteq\max_{\{\pi_i,\rho_i\}}\chi(\{\pi_i,\Phi(\rho_i)\})=\max_{\rho\in\S(\H_A)}\bar{C}(\Phi,\rho),
\end{equation}
where the first maximum is over all ensembles of states in $\S(\H_A)$.

An ensemble $\{\pi_i,\rho_i\}$, at which the first maximum in (\ref{chi-cap}) is attained, is called
\emph{optimal } for the channel $\,\Phi$ \cite{Sch-West}.

By the HSW theorem the classical  capacity of the channel $\Phi$ can
be expressed by the following regularization formula
$$
C(\Phi)=\lim_{n\rightarrow+\infty} n^{-1}\bar{C}(\Phi^{\otimes n}).
$$

By the BSST theorem the entanglement-assisted classical capacity of the
channel $\Phi$ is determined as follows
\begin{equation}\label{ea-cap}
C_{\mathrm{ea}}(\Phi)=\max_{\rho\in\S(\H_A)}I(\Phi,\rho).
\end{equation}

According to \cite{Sh-16} introduce the following
notion. \medskip
\begin{definition}\label{ef-subch} \emph{Let $\H^{\chi}_{\Phi}$ be the minimal subspace of $\H_A$ containing elements of
all optimal ensembles for the channel
$\,\Phi:\S(\H_A)\rightarrow\S(\H_B)$. The restriction $\Phi_{\chi}$
of the channel $\,\Phi$ to the set $\S(\H^{\chi}_{\Phi})$ is called the
$\chi$\nobreakdash-\hspace{0pt}essential part (subchannel) of the channel $\,\Phi$.}
\end{definition}\medskip

It is clear that $\,\H^{\chi}_{\Phi}=\H_A$ means existence of
an optimal ensemble for the channel
$\Phi$ with the full rank average state.
If $\,\H^{\chi}_{\Phi}\neq\H_A$ then pure states corresponding to vectors in
$\H_A\setminus\H^{\chi}_{\Phi}$ are useless for non-entangled coding of classical information and
hence it is natural to consider the $\chi$\nobreakdash-\hspace{0pt}essential subchannel
$\Phi_{\chi}$ instead of the channel $\,\Phi$ dealing with the
Holevo capacity of the channel $\Phi$ (which coincides with the classical capacity if $C_{\mathrm{ea}}(\Phi)=\bar{C}(\Phi)$).

By definition $\bar{C}(\Phi_{\chi})=\bar{C}(\Phi)$. Hence
$C_{\mathrm{ea}}(\Phi)=\bar{C}(\Phi)$ implies that \break
$C_{\mathrm{ea}}(\Phi_{\chi})=C_{\mathrm{ea}}(\Phi)$. Thus, in this case speaking about the entanglement-assisted capacity of the
channel $\Phi$ we  may also consider the $\chi$\nobreakdash-\hspace{0pt}essential subchannel
$\Phi_{\chi}$ instead of the channel $\,\Phi$.\medskip

Now we can formulate our main result.\medskip

\begin{theorem}\label{c-q-ch}
\emph{Let $\,\Phi:\S(\H_A)\rightarrow\S(\H_B)$ be a quantum channel.}
\medskip

A) \emph{If $\,\Phi$
is a c-q channel then $\,C_{\mathrm{ea}}(\Phi)=\bar{C}(\Phi)$;} \footnote{This assertion seems well known. I would be grateful for the reference.} \medskip

B) \emph{If $\,C_{\mathrm{ea}}(\Phi)=\bar{C}(\Phi)$ then the $\chi$-essential part of $\,\Phi$
is a c-q channel.}
\end{theorem}\medskip

\begin{remark}\label{c-q-ch-r} The presence of "the $\chi$-essential part of $\Phi$"
in assertion B is natural, since the remarks before Theorem \ref{c-q-ch} show that the equality $C_{\mathrm{ea}}(\Phi)=\bar{C}(\Phi)$ can not give information about action of the channel $\Phi$ on states  not supported by the subspace $\H^{\chi}_{\Phi}$. This conclusion is confirmed by the example of non-entanglement-breaking channel $\Phi$ such that
$C_{\mathrm{ea}}(\Phi)=\bar{C}(\Phi)$ proposed in \cite{BSST+} and mathematically described in \cite[Example 3]{Sh-16}.
\end{remark}
\medskip

\begin{corollary}\label{c-q-ch-c}
\emph{Let $\,\Phi:\S(\H_A)\rightarrow\S(\H_B)$ be a quantum channel
for which there exists an optimal ensemble with the full rank
average state. Then}
$$
C_{\mathrm{ea}}(\Phi)=\bar{C}(\Phi)\quad\Leftrightarrow\quad\Phi
\;\,\textit{is a c-q channel}.
$$
\end{corollary}\medskip

\textbf{Proof of Theorem \ref{c-q-ch}.} A)  If the channel $\Phi$ has representation (\ref{c-q-rep}) then $\Phi=\Phi\circ\Pi$,
where $\Pi(\rho)=\sum_{k=1}^{\dim\H_A}\langle k|\rho|k\rangle|k\rangle\langle
k|$. By the chain rule for the quantum mutual information we have
\begin{equation*}
I(\Phi,\rho)=I(\Phi\circ\Pi,\rho)\leq I(\Phi,\Pi(\rho)).
\end{equation*}
Hence the supremum in expression (\ref{ea-cap}) can be taken only over states diagonizable in the basis $\{|k\rangle\}$. So, to prove the equality
$C_{\mathrm{ea}}(\Phi)=\bar{C}(\Phi)$ it suffices to show that $I(\Phi,\rho)=\bar{C}(\Phi,\rho)$ for any such state $\rho$.

Since the channel $\Phi$ has representation (\ref{c-q-rep}), Proposition 1 in \cite{Sh-18} implies that the channel $\widehat{\Phi}$ is reversible with respect to the family $\{|k\rangle\langle
k|\}$ of pure states (i.e. there exists a channel $\Psi$ such that $\Psi(\widehat{\Phi}(|k\rangle\langle
k|))=|k\rangle\langle
k|$ for all $k$). It follows that $\bar{C}(\widehat{\Phi},\rho)=H(\rho)$ for any state $\rho$ diagonizable in the basis $\{|k\rangle\}$. By expression (\ref{mi-rep++}) the last equality is equivalent to the equality $I(\Phi,\rho)=\bar{C}(\Phi,\rho)$.

\smallskip

B) By replacing the channel $\Phi$ by its $\chi$-essential subchannel we may assume that
there exists an optimal ensemble $\{\pi_i,\rho_i\}$ of pure states for the channel $\Phi$ with the full rank
average $\bar{\rho}$. By expression (\ref{mi-rep++}) we have
\begin{equation}\label{chain}
C_{\mathrm{ea}}(\Phi)=\bar{C}(\Phi)\quad\Rightarrow\quad I(\Phi,\bar{\rho})=\bar{C}(\Phi,\bar{\rho})\quad\Leftrightarrow\quad \bar{C}(\widehat{\Phi},\bar{\rho})=H(\bar{\rho}).
\end{equation}
Since $H(\bar{\rho})=\chi(\{\pi_i,\rho_i\})$ and $\bar{C}(\widehat{\Phi},\bar{\rho})=\chi(\{\pi_i,\widehat{\Phi}(\rho_i)\})$ (this can be shown by using (\ref{chi-fun+}) and coincidence of the functions $\hat{H}_{\Phi}$ and $\hat{H}_{\widehat{\Phi}}$), the last equality in (\ref{chain}) and Theorem 5 in \cite{Sh-18} show that $\Phi$ is a c-q channel (it coincides with the complementary channel to the channel $\widehat{\Phi}$ up to isometrical equivalence). $\square$

Theorem \ref{c-q-ch} makes it possible to strengthen Proposition 4 in \cite{Sh-16} as follows.

\smallskip

\begin{property}\label{degradable}
\emph{Let $\,\Phi:\S(\H_A)\rightarrow\S(\H_B)$ be a degradable
channel. Then the following statements are equivalent:}
\begin{enumerate}[(i)]
  \item $C_{\mathrm{ea}}(\Phi)=\bar{C}(\Phi);$
  \item $C_{\mathrm{ea}}(\Phi)=\bar{C}(\Phi)=\log\dim\H_A;$
  \item \emph{$\Phi$ is a c-q channel having representation (\ref{c-q-rep}) in which $\{\sigma_k\}$ is a collection of states in $\H_B$ with pairwise orthogonal supports.}
\end{enumerate}
\end{property}
\smallskip
\textbf{Proof.} $\mathrm{(i)\Rightarrow(ii)}\,$ follows from  Proposition 4 in \cite{Sh-16}.

\smallskip

$\mathrm{(ii)\Rightarrow(iii)}$.  $\,\bar{C}(\Phi)=\log\dim\H_A\,$ implies $\,\Phi_{\chi}=\Phi\,$ and hence Theorem \ref{c-q-ch}B shows that $\Phi$ is a c-q channel.

Assume that the states of the collection $\{\sigma_k\}$ from representation (\ref{c-q-rep}) of the channel $\Phi$ are decomposed as follows  $\;\sigma_k=\sum_{i=1}^{\dim\H_B}|\psi_{ki}\rangle\langle
\psi_{ki}|\;$. Then $\Phi(\rho)=\sum_{k,i}W_{ki}\rho\, W^*_{ki}$,
where $W_{ki}= |\psi_{ki}\rangle\langle k|$ ($\{|k\rangle\}$ is the basis from representation (\ref{c-q-rep})), and by using the standard representation of the complementary channel (see \cite[formula (11)]{H-comp-ch}) we obtain
\begin{equation}\label{c-ch-rep}
\widehat{\Phi}(\rho)=\sum_{k,\,l=1}^{\dim\H_A}\langle
k|\rho|l\rangle|k\rangle\langle l|
\otimes\sum_{i,\,j=1}^{\dim\H_B}\langle\psi_{lj}|\psi_{ki}\rangle|i\rangle\langle
j|\in\S(\H_A\otimes\H_B).
\end{equation}
Since $\Phi$ is a degradable channel having  representation (\ref{c-q-rep}), we have\break $\widehat{\Phi}(|k\rangle\langle
l|)=\Psi\circ\Phi(|k\rangle\langle l|)=0$ for all $k\neq l$.
Hence (\ref{c-ch-rep}) shows that $\langle\psi_{lj}|\psi_{ki}\rangle=0$ for all  $i,j$ and all $k\neq l$, which means that
$\mathrm{supp}\sigma_k\perp\mathrm{supp}\sigma_l$ for all $k\neq
l$.

\smallskip

$\mathrm{(iii)\Rightarrow(i)}\,$ follows from Theorem \ref{c-q-ch}A.
$\square$
\medskip

\section{Infinite dimensional channels}

\subsection{On coincidence of the constrained Holevo capacity and the quantum mutual information}

Let $\,\Phi:\S(\H_A)\rightarrow\S(\H_B)$ be an arbitrary finite or infinite dimensional quantum channel.
In this subsection we consider conditions for the equality in the general inequality
\begin{equation*}
\bar{C}(\Phi,\rho)\leq I(\Phi,\rho),\quad \rho\in\S(\H_A)
\end{equation*}
This inequality can be proved by using expression (\ref{mi-rep++})
valid under the condition $H(\rho)<+\infty$ and a simple approximation.
\smallskip

We have to introduce some additional notions.
\smallskip

A continuous (generalized) ensemble of quantum states can be defined
as a Borel probability measure $\mu$ on the set $\S(\H)$. The $\chi$-quantity of
such ensemble (measure) $\mu$ is defined as follows (cf.
\cite{H-Sh-2})
\begin{equation}\label{h-q-c}
\chi(\mu)=\int_{\mathfrak{S}(\mathcal{H})}H(\rho\Vert\bar{\rho}(\mu))\mu(d\rho),
\end{equation}
where $\bar{\rho}(\mu)$ is the barycenter of the measure $\mu$
defined by the Bochner integral
$$
\bar{\rho}(\mu)=\int_{\mathfrak{S}(\mathcal{H})}\rho \mu(d\rho ).
$$
If $H(\bar{\rho}(\mu))<+\infty$ then
$\chi(\mu)=H(\bar{\rho}(\mu))-\int_{\mathfrak{S}(\mathcal{H})}H(\rho)\mu
(d\rho)$ \cite{H-Sh-2}.\smallskip

Denote by $\P(\mathcal{A})$ the set of all Borel probability
measures on a closed subset $\mathcal{A}\subset\S(\H)$ endowed
with the weak convergence topology \cite{Par}.\smallskip

The image of a continuous ensemble $\mu\in\P(\S(\H_A))$ under a
channel $\Phi:\S(\H_A)\rightarrow\S(\H_B)$ is a continuous ensemble
corresponding to the measure
$\Phi(\mu)\doteq\mu\circ\Phi^{-1}\in\P(\S(\H_B))$. Its $\chi$-quantity can be expressed as follows
\begin{equation}\label{chi-phi-mu}
\begin{array}{c}
\displaystyle\chi(\Phi(\mu))\doteq\int_{\mathfrak{S}(\H_A)}H(\Phi(\rho
)\Vert \Phi (\bar{\rho}(\mu)))\mu(d\rho )\\\displaystyle=H(\Phi
(\bar{\rho}(\mu)))-\int_{\mathfrak{S}(\H_A)}H(\Phi(\rho))\mu
(d\rho),
\end{array}
\end{equation}
where the second formula is valid under the condition $H(\Phi
(\bar{\rho}(\mu)))<+\infty$.\medskip

The constrained Holevo capacity defined by (\ref{chi-fun}) can be also expressed as
follows
\begin{equation}\label{chi-fun++}
\bar{C}(\Phi,\rho)=\sup_{\mu\in\P(\S(\H_A)),\,
\bar{\rho}(\mu)=\rho}\chi(\Phi(\mu)).
\end{equation}
This expression follows from Corollary 1 in
\cite{H-Sh-2} with $\mathcal{A}=\{\rho\}$ (where the
constrained Holevo capacity is denoted
$\chi_{\Phi}(\rho)$).
\medskip

\begin{theorem}\label{inf-dim} \emph{Let $\,\Phi:\S(\H_A)\rightarrow\S(\H_B)$ be a quantum channel
and $\rho$ be a state in $\,\S(\H_A)$ with the support $\H_{\rho}$.}\smallskip

A) \emph{If the restriction $\Phi|_{\S(\H_{\rho})}$ of the channel $\Phi$ to the set $\S(\H_{\rho})$ is a c\nobreakdash-\hspace{0pt}q channel
 having  representation (\ref{c-q-rep}) with $\H_A=\H_{\rho}$, in which $\{|k\rangle\}$ is an orthonormal basis of eigenvectors of the state
 $\rho$, then $\,\bar{C}(\Phi,\rho)=I(\Phi,\rho)\leq+\infty$}.\medskip

B) \emph{If $H(\rho)<+\infty$ and the following condition holds
\begin{equation}\label{a-cond}
\exists\;\mu\in\P(\S(\H_A))\;\;\text{such that}\;\;
\bar{\rho}(\mu)=\rho\;\; and\;\; \bar{C}(\Phi,\rho)=\chi(\Phi(\mu)),
\end{equation}
which means that the supremum in
(\ref{chi-fun++}) is attainable, then}
$$
\bar{C}(\Phi,\rho)=I(\Phi,\rho)<+\infty\quad\Rightarrow\quad\Phi|_{\S(\H_{\rho})}\;\, \textit{is a c-q channel}.
$$

\emph{Condition (\ref{a-cond}) is valid if either
$H(\Phi(\rho))<+\infty$ or one of the functions $\,\sigma\mapsto
H(\Phi(\sigma)\|\Phi(\rho))$, $\,\sigma\mapsto
H(\widehat{\Phi}(\sigma)\|\widehat{\Phi}(\rho))\,$ is continuous and bounded on the set
$\,\mathrm{extr}\S(\H_A)$.}
\end{theorem}\medskip

\begin{remark}\label{inf-dim-r}
Example 5 in \cite{Sh-16} shows that the assertion of Theorem \ref{inf-dim-r}A is not valid if the state $\rho$ is not diagonizable in the basis     $\{|k\rangle\}$.
\end{remark}\medskip

\textbf{Proof.} A) We may assume that $\H_{\rho}=\H_A$. Let $\rho=\sum_{k=1}^{\dim\H_A}\lambda_k|k\rangle\langle k|$ and $\rho_n=[\sum_{k=1}^{n}\lambda_k]^{-1}\sum_{k=1}^{n}\lambda_k|k\rangle\langle k|$. Then the sequence $\{\rho_n\}$ converges to the state $\rho$ and $H(\rho_n)<+\infty$ for all $n$.

If the channel $\Phi$ has representation (\ref{c-q-rep}) then the channel $\widehat{\Phi}$ is reversible with respect to the family $\{|k\rangle\langle k|\}_{k=1}^{\dim\H_A}$ by Proposition 1 in \cite{Sh-18}. Hence  $\bar{C}(\widehat{\Phi},\rho_n)=H(\rho_n)$ and expression (\ref{mi-rep++}) implies $\bar{C}(\Phi,\rho_n)=I(\Phi,\rho_n)$ for all $n$. It  follows that $\bar{C}(\Phi,\rho)=I(\Phi,\rho)$, since by using concavity and lower semicontinuity of the nonnegative functions
$\rho\mapsto\bar{C}(\Phi,\rho)$ and $\rho\mapsto I(\Phi,\rho)$ it is easy to show that
$$
\lim_{n\rightarrow+\infty}\bar{C}(\Phi,\rho_n)=\bar{C}(\Phi,\rho)\leq+\infty\quad \textup{and} \quad\lim_{n\rightarrow+\infty}I(\Phi,\rho_n)=I(\Phi,\rho)\leq+\infty.
$$

B) Without loss of generality we may consider that the measure $\mu$ in
(\ref{a-cond}) belongs to the set $\P(\mathrm{extr}\S(\H_A))$. This follows from convexity of the function
$\sigma\mapsto
H(\Phi(\sigma)\|\Phi(\rho))$, since for an arbitrary measure $\mu\in\P(\S(\H_A))$ there exists a measure $\hat{\mu}\in\P(\mathrm{extr}\S(\H_A))$ such that $\bar{\rho}(\hat{\mu})=\bar{\rho}(\mu)$ and $\int f(\sigma)\hat{\mu}(d\sigma)\geq\int f(\sigma)\mu(d\sigma)$ for any convex lower semicontinuous nonnegative function $f$ on the set $\S(\H_A)$ (this measure $\hat{\mu}$ can be constructed by using the arguments from the proof of the Theorem in \cite{H-Sh-2}).

By expression (\ref{mi-rep++}) the equality $\bar{C}(\Phi,\rho)=I(\Phi,\rho)$ is equivalent to the equality
$H(\rho)=\bar{C}(\widehat{\Phi},\rho)$. By the remark after Lemma
\ref{inf-dim-l} in the Appendix condition (\ref{a-cond}) implies
that $\bar{C}(\widehat{\Phi},\rho)=\chi(\widehat{\Phi}(\mu))$. Since
$H(\rho)=\chi(\mu)$, the equality $H(\rho)=\bar{C}(\widehat{\Phi},\rho)$
shows that the channel $\widehat{\Phi}$ preserves the $\chi$-quantity of the measure $\mu$. By Theorem 5 in \cite{Sh-18} the
restriction to the
set $\S(\H_{\rho})$ of the complementary channel to the channel $\widehat{\Phi}$  is a c-q channel.\smallskip

If $H(\Phi(\rho))<+\infty$ then  condition (\ref{a-cond}) holds by
Corollary 2 in \cite{H-Sh-2}.

If the function $\sigma\mapsto H(\Phi(\sigma)\|\Phi(\rho))$ is
continuous and bounded on the set $\,\mathrm{extr}\S(\H_A)$ then the
function $\P(\mathrm{extr}\S(\H_A))\ni\mu\mapsto\chi(\Phi(\mu))$ is
continuous by the definition of the weak convergence. Since the
subset of $\P(\mathrm{extr}\S(\H_A))$ consisting of measures with
the barycenter $\rho$ is compact by Proposition 2 in \cite{H-Sh-2},
the last function attains its least upper bound on this subset.

If the function $\sigma\mapsto H(\widehat{\Phi}(\sigma)\|\widehat{\Phi}(\rho))$ is
continuous and bounded on the set $\,\mathrm{extr}\S(\H_A)$
then the similar arguments show attainability of the supremum in the definition of the value $\bar{C}(\widehat{\Phi},\rho)$, which is equivalent to (\ref{a-cond}) by the remark after Lemma
\ref{inf-dim-l} in the Appendix. $\square$
\smallskip

\subsection{Conditions for the equality $C_{\mathrm{ea}}(\Phi)=\bar{C}(\Phi)$ for channels with linear constraints}

In this subsection we derive from Theorem
\ref{inf-dim} the necessary and sufficient conditions for coincidence of the entanglement-assisted classical capacity and the Holevo
capacity of an infinite dimensional
channel $\Phi:\S(\H_A)\rightarrow\S(\H_B)$ with the constraint defined by the inequality
\begin{equation}\label{lc}
 \Tr H\rho\leq h,\quad h>0,
\end{equation}
where $H$ is a positive operator in $\H_A$ -- Hamiltonian of the input quantum
system.\footnote{Speaking about capacities of infinite dimensional quantum
channels we have to impose particular constraints on the choice of
input code-states to avoid infinite values of the capacities and to
be consistent with the physical implementation of the process of
information transmission \cite{H-c-w-c}.} The operational
definitions of the unassisted and the
entanglement-assisted classical capacities of a quantum channel with
constraint (\ref{lc}) are given in \cite{H-c-w-c}, where the
corresponding generalizations of the HSW and BSST theorems are
proved. \medskip

The Holevo capacity of the channel $\Phi$ with constraint (\ref{lc})
can be defined as follows
\begin{equation}\label{chi-cap+}
\bar{C}(\Phi|H,h)=\sup_{\mu\in\P(\S(\H_A)),\Tr H\bar{\rho}(\mu)\leq h}\chi(\Phi(\mu)),
\end{equation}
where the supremum can be taken over all measures in $\P(\S(\H_A))$ supported by pure states or, equivalently,
\begin{equation}\label{chi-cap++}
\bar{C}(\Phi|H,h)=\sup_{\rho\in\S(\H_A),\,\Tr H\rho\leq h}\bar{C}(\Phi,\rho).
\end{equation}

If the supremum in (\ref{chi-cap+}) is achieved at a measure  $\mu_*$ then this measure is called optimal for the channel $\Phi$ with constraint
(\ref{lc}). The sufficient condition for existence of optimal measures and examples showing that optimal measures do not exist in general can be found in \cite{H-Sh-2}.

An optimal measure for the channel $\Phi$ with constraint
(\ref{lc}) exists if and only if the supremum in
(\ref{chi-cap++}) is achieved at a state $\rho$ for which the condition (\ref{a-cond}) holds (the sufficient conditions for this
are presented at the end of Theorem \ref{inf-dim}).

By the generalized HSW theorem (\cite[Proposition 3]{H-c-w-c}) the
classical capacity of the channel $\Phi$ with constraint (\ref{lc})
can be expressed by the following regularization formula
$$
C(\Phi|H,h)=\lim_{n\rightarrow+\infty} n^{-1}\bar{C}(\Phi^{\otimes
n}|H_n, nh),
$$
where $H_n=H\otimes I\otimes...\otimes I + I\otimes H\otimes
I\otimes...\otimes I + ... + I\otimes...\otimes I\otimes H$ (each of
$n$ summands consists of $n$ multiples).

By the generalized BSST theorem (\cite[Proposition 4]{H-c-w-c}) the
entanglement-assisted classical capacity of the channel $\Phi$  with
constraint (\ref{lc}) is determined as follows
\begin{equation}\label{ea-cap+}
C_{\mathrm{ea}}(\Phi|H,h)=\sup_{\rho\in\S(\H_A),\,\Tr H\rho\leq h}I(\Phi, \rho).
\end{equation}
This expression is proved in \cite{H-c-w-c} under the particular
technical conditions on the channel $\Phi$ and the operator
$H$, which can be removed by using the approximation method \cite{Sh-H}. We will assume that expression (\ref{ea-cap+}) is valid.\medskip

Theorem \ref{inf-dim} implies the following conditions
for coincidence of $\bar{C}(\Phi|H,h)$ and
$C_{\mathrm{ea}}(\Phi|H,h)$.
\smallskip

\begin{corollary}\label{inf-dim+} \emph{Let $\Phi:\S(\H_A)\rightarrow\S(\H_B)$ be a quantum
channel and $H$ be a positive operator in $\,\S(\H_A)$.}\smallskip

A) \emph{If $\,\Phi$ is a c-q channel having representation (\ref{c-q-rep}) and the operator $H$ is diagonizable\footnote{This means that any spectral projector of the operator $H$ is diagonizable
in the basis $\{|k\rangle\}$.}
in the basis $\{|k\rangle\}$ in (\ref{c-q-rep}) then $\,\bar{C}(\Phi|H,h)=C_{\mathrm{ea}}(\Phi|H,h)$;}
\smallskip

B) \emph{If $\,\bar{C}(\Phi|H,h)=C_{\mathrm{ea}}(\Phi|H,h)<+\infty$ and
the supremum in (\ref{chi-cap+}) is achieved at a measure  $\mu_*$
such that $H(\bar{\rho}(\mu_*))<+\infty\,$  then
the restriction of the channel $\,\Phi$ to the set $\,\S(\H_{\bar{\rho}(\mu_*)})$,
$\,\H_{\bar{\rho}(\mu_*)}=\mathrm{supp}\bar{\rho}(\mu_*)$, is a c-q channel.}\smallskip

\emph{The condition concerning existence of the measure $\mu_*$ holds if the subset of $\,\S(\H_A)$ defined by inequality (\ref{lc}) is compact and the output entropy of the channel $\Phi$ (the function $\rho\mapsto H(\Phi(\rho))$) is continuous on
this subset.}
\end{corollary}\medskip

\begin{remark}\label{inf-dim-c-r}
Example 5 in \cite{Sh-16} shows that the assertion of Corollary \ref{inf-dim+}A is not valid if the operator $H$ is not diagonizable in the basis     $\{|k\rangle\}$.
\end{remark}\medskip

\textbf{Proof.} A) If the channel $\Phi$ has representation (\ref{c-q-rep}) then $\Phi=\Phi\circ\Pi$,
where $\Pi(\rho)=\sum_{k=1}^{\dim\H_A}\langle k|\rho|k\rangle|k\rangle\langle
k|$. By the chain rule for the quantum mutual information we have
\begin{equation}\label{a-ineq}
I(\Phi,\rho)=I(\Phi\circ\Pi,\rho)\leq I(\Phi,\Pi(\rho)).
\end{equation}
If the operator $H$ is diagonizable in the basis $\{|k\rangle\}$ then the inequality\break $\Tr H\rho\leq h$ implies the inequality
$\Tr H\Pi(\rho)\leq h$. Hence (\ref{a-ineq}) shows that the supremum in expression (\ref{ea-cap+}) can be taken only over states diagonizable in the basis $\{|k\rangle\}$. Since $\,\bar{C}(\Phi,\rho)=I(\Phi,\rho)$ for any such state $\rho$ by Theorem \ref{inf-dim}A, we have $\,\bar{C}(\Phi|H,h)=C_{\mathrm{ea}}(\Phi|H,h)$.
\medskip

B) The main assertion immediately follows from Theorem \ref{inf-dim}A, since its condition implies
$\,\bar{C}(\Phi,\bar{\rho}(\mu_*))=I(\Phi,\bar{\rho}(\mu_*))$. The sufficient condition for existence of the measure $\mu_*$ follows from the Theorem in \cite{H-Sh-2}. $\square$

\medskip
\textbf{Example.} The condition for existence of an optimal measure in Corollary \ref{inf-dim+}B holds for a
Gaussian channel $\Phi$ with the power constraint of the form
(\ref{lc}), where $H=R^{T}\epsilon R$ is the many\nobreakdash-\hspace{0pt}mode oscillator
Hamiltonian (see the remark after Proposition 3 in \cite{H-Sh-2}). So, if we assume that $\bar{\rho}(\mu_*)$ is a Gaussian state, then Corollary \ref{inf-dim+}B shows that $\bar{C}(\Phi|H,h)=C_{\mathrm{ea}}(\Phi|H,h)$ may be valid only if  $\Phi$ is a c-q channel. \medskip

The above assumption holds provided the conjecture of Gaussian optimizers is valid for the channel $\Phi$ (see \cite{E&W,G-opt} and the references therein).

\section*{Appendix}

Let $\,\Phi:\S(\H_A)\rightarrow\S(\H_B)$ be a quantum channel and
$\,\widehat{\Phi}:\S(\H_A)\rightarrow\S(\H_E)$ be its complementary
channel. In finite dimensions the \emph{coherent information} of the
channel $\Phi$ at any state $\rho$ can be defined as a difference
between $H(\Phi(\rho))$ and  $H(\widehat{\Phi}(\rho))$ \cite{N&Ch,Sch}.
Since in infinite dimensions these values may be infinite even for
the state $\rho$ with finite entropy, for any such state the
coherent information can be defined via the quantum mutual information as follows
$$
I_c(\Phi,\rho)=I(\Phi,\rho)-H(\rho).
$$

Let $\rho$ be a state in $\,\S(\H_A)$ with finite entropy. By
monotonicity of the relative entropy the values $\chi(\Phi(\mu))$ and
$\chi(\widehat{\Phi}(\mu))$ do not exceed $H(\rho)=\chi(\mu)$ for
any measure $\mu\in\P(\mathrm{extr}\S(\H_A))$ with the barycenter
$\rho$. The following lemma can be considered as a
generalized version of the observation in
\cite{Sch}.
\smallskip
\begin{lemma}\label{inf-dim-l}
\emph{Let $\mu$ be a measure in $\P(\mathrm{extr}\S(\H_A))$ with the
barycenter $\rho$. Then
\begin{equation}\label{chi-dif}
\chi(\Phi(\mu))-\chi(\widehat{\Phi}(\mu))=I(\Phi,\rho)-H(\rho)=I_c(\Phi,\rho).
\end{equation}}
\end{lemma}

This lemma shows, in particular, that the difference
$\chi(\Phi(\mu))-\chi(\widehat{\Phi}(\mu))$ does not depend on
$\mu$. So, if the supremum in expression
(\ref{chi-fun++}) for the value $\bar{C}(\Phi,\rho)$ is achieved at
some measure $\mu_*$ then the supremum in the similar  expression
for the value $\bar{C}(\widehat{\Phi},\rho)$ is achieved at this
measure $\mu_*$ and vice versa.

\smallskip

\textbf{Proof.} If $H(\Phi(\rho))<+\infty$ then
$H(\widehat{\Phi}(\rho))<+\infty$ by the triangle inequality and
(\ref{chi-dif}) can be derived from (\ref{mi-rep+}) by using  the
second formula in (\ref{chi-phi-mu}) and  by noting that the
functions $\rho\mapsto H(\Phi(\rho))$ and $\rho\mapsto
H(\widehat{\Phi}(\rho))$ coincide on the set of pure states. In
general case it is necessary to use the approximation method to
prove (\ref{chi-dif}). To realize this method we have to introduce
some additional notions. \smallskip

Let
$\mathfrak{T}_{1}(\mathcal{H})=\{A\in\mathfrak{T}(\mathcal{H})\,|\,A\geq
0,\;\Tr A\leq 1\}$. We will use the  following two extensions of the
von Neumann entropy to the set $\mathfrak{T}_{1}(\mathcal{H})$
(cf.\cite{L})
$$
S(A)=-\Tr A\log A\quad\textup{and}\quad H(A)=S(A)+\Tr A\log \Tr
A,\quad \forall A \in\mathfrak{T}_{1}(\mathcal{H}).
$$
Nonnegativity, concavity and lower semicontinuity of the von Neumann
entropy imply the same properties of the functions $S$ and $H$ on
the set $\mathfrak{T}_{1}(\mathcal{H})$.

The relative entropy for two operators $A$ and $B$ in
$\mathfrak{T}_{1}(\mathcal{H})$ is defined as follows (cf.\cite{L})
$$
H(A\,\|B)=\sum_{i}\langle i|\,(A\log A-A\log B+B-A)\,|i\rangle,
$$
where $\{|i\rangle\}$ is the orthonormal basis of eigenvectors of
$A$. By means of this extension of the relative entropy the $\chi$-quantity of a measure $\mu$ in $\P(\mathfrak{T}_{1}(\H))$ is defined
by expression (\ref{h-q-c}).

A completely positive trace-non-increasing linear map
$\Phi:\mathfrak{T}(\H_A)\rightarrow\mathfrak{T}(\H_B)$ is called
\emph{quantum operation} \cite{N&Ch}.  For any quantum operation $\Phi$ the
Stinespring representation (\ref{Stinespring-rep}) holds, in which
$V$ is a contraction. The complementary operation
$\widehat{\Phi}:\mathfrak{T}(\H_A)\rightarrow\mathfrak{T}(\H_E)$ is defined
via this representation by (\ref{c-channel}).

By the obvious modification of the arguments used in the proof of
Proposition 1 in \cite{H-Sh-2} one can show that the function
$\mu\mapsto\chi(\mu)$ is lower semicontinuous on the set
$\P(\mathfrak{T}_{1}(\H))$ and that for an arbitrary quantum
operation $\Phi$ and a measure $\mu\in\P(\S(\H_A))$ such that
$S(\Phi(\bar{\rho}(\mu)))<+\infty$ the $\chi$-quantity of the measure
$\Phi(\mu)\doteq\mu\circ\Phi^{-1}\in\P(\mathfrak{T}_{1}(\H_B))$ can
be expressed as follows
\begin{equation}\label{formula}
\chi(\Phi(\mu))=
S(\Phi(\bar{\rho}(\mu))-\int_{\S(\H_A)}S(\Phi(\rho))\mu(d\rho).
\end{equation}

We are now in a position to prove (\ref{chi-dif}) in general case.

Note that for a given measure
$\mu\in\P(\S(\H_A))$ the function
$\Phi\mapsto\chi(\Phi(\mu))$ is lower semicontinuous on the set of
all quantum operations endowed with the strong convergence topology
(in which $\Phi_n\rightarrow\Phi$ means
$\Phi_n(\rho)\rightarrow\Phi(\rho)$ for all $\rho$ in the trace norm \cite{Sh-H}). This follows
from lower semicontinuity of the function $\mu\mapsto\chi(\mu)$ on
the set $\P(\mathfrak{T}_1(\H_B))$, since for an arbitrary sequence
$\{\Phi_n\}$ of quantum operations strongly converging to a quantum
operation $\Phi$ the sequence
$\{\Phi_n(\mu)\}\subset\P(\mathfrak{T}_1(\H_B))$ weakly converges to
the measure $\Phi(\mu)$ (this can be verified directly by using the
definition of the weak convergence and by noting that for sequences
of quantum operations the strong convergence is equivalent to the
uniform convergence on compact subsets of $\S(\H_A)$).

Let $\{P_n\}$ be an increasing sequence of finite rank projectors in
$\B(\H_B)$ strongly converging to $I_B$. Consider the sequence of
quantum operations $\Phi_n=\Pi_n\circ\Phi$, where
$\Pi_n(\cdot)=P_n(\cdot)P_n$. Then
\begin{equation}\label{c-oper}
    \widehat{\Phi}_n(\rho)=\Tr_{\H_B}P_n\otimes I_{\H_E}V\rho V^*,\quad
    \rho\in\S(\H_A),
\end{equation}
where $V$ is the isometry from Stinespring representation
(\ref{Stinespring-rep}) for the channel $\Phi$.

The sequences $\{\Phi_n\}$ and
$\{\widehat{\Phi}_n\}$ strongly converges to the channels $\Phi$ and
$\widehat{\Phi}$ correspondingly. Let
$\rho=\sum_{k}\lambda_k|k\rangle\langle k|$ and
$|\varphi_{\rho}\rangle=\sum_{k}\sqrt{\lambda_k}|k\rangle\otimes|k\rangle$.
Since $S(\Phi_n(\rho))<+\infty$, the triangle inequality implies
$S(\widehat{\Phi}_n(\rho))<+\infty$. So, we have
\begin{equation}\label{I-n}
\begin{array}{c}
\displaystyle I(\Phi_n,\rho)=H\left(\Phi_n \otimes \id_{R}
(|\varphi_{\rho}\rangle\langle\varphi_{\rho}|)\, \|\, \Phi_n(\rho)
\otimes \varrho\right)\\\\ \displaystyle =
-S(\widehat{\Phi}_n(\rho))+S(\Phi_n(\rho))+a_n=-\chi(\widehat{\Phi}_n(\mu))+\chi(\Phi_n(\mu))+a_n,
\end{array}
\end{equation}
where $a_n=-\sum_{k}\Tr(\Phi_n(|k\rangle\langle
k|))\lambda_k\log\lambda_k$ and the last equality is obtained by
using (\ref{formula}) and coincidence of the functions $\rho\mapsto
S(\Phi(\rho))$ and $\rho\mapsto S(\widehat{\Phi}(\rho))$ on the set
of pure states.

Since the function $\Phi\mapsto I(\Phi,\rho)$ is lower
semicontinuous (by lower semicontinuity of the relative entropy) and
$I(\Phi_n,\rho)\leq I(\Phi,\rho)$  for all $n$ by monotonicity
of the relative entropy under action the quantum operation
$\Pi_n\otimes\id_{R}$, we have
\begin{equation}\label{lim-r-1}
    \lim_{n\rightarrow+\infty}I(\Phi_n,\rho)=I(\Phi,\rho).
\end{equation}

We will also show that
\begin{equation}\label{lim-r-2}
\lim_{n\rightarrow+\infty} \chi(\Phi_n(\mu))= \chi(\Phi(\mu))\quad
\textrm{and}\quad
\lim_{n\rightarrow+\infty}\chi(\widehat{\Phi}_n(\mu))=
\chi(\widehat{\Phi}(\mu)).
\end{equation}

The first relation in (\ref{lim-r-2}) follows from lower
semicontinuity of the function $\Phi\mapsto\chi(\Phi(\mu))$
(established before) and the inequality $\chi(\Phi_n(\mu))\leq
\chi(\Phi(\mu))$ valid for all $n$ by monotonicity of the $\chi$-quantity
under action of the quantum operation $\Pi_n$.

To prove the second  relation in (\ref{lim-r-2}) note that
(\ref{c-oper}) implies
$\widehat{\Phi}_n(\rho)\leq\widehat{\Phi}(\rho)$ for any state
$\rho\in \S(\H_A)$. Thus Lemma 2 in \cite{Sh-H} shows that
\begin{equation}\label{chi-ineq}
\chi(\widehat{\Phi}_{n}(\mu))\leq \chi(\widehat{\Phi}(\mu))+
f(\Tr\widehat{\Phi}_{n}(\rho))
\end{equation}
where $f(x)=-2x\log x-(1-x)\log(1-x)$, for any measure
$\mu\in\P(\S(\H_A))$ with finite support and the barycenter $\rho$.
To prove that (\ref{chi-ineq}) holds for any measure
$\mu\in\P(\S(\H_A))$ with the barycenter $\rho$ one can
take the sequence $\{\mu_n\}$ of measures with finite support and
the barycenter $\rho$ constructed in the proof of Lemma 1 in
\cite{H-Sh-2}, which weakly converges to the measure $\mu$, and use
lower semicontinuity of the function $\mu\mapsto\chi(\Psi(\mu))$,
where $\Psi$ is a quantum operation,  and the inequality
$\chi(\widehat{\Phi}(\mu_n))\leq\chi(\widehat{\Phi}(\mu))$ valid for
all $n$ by the construction of the sequence $\{\mu_n\}$ and
convexity of the relative entropy.

Inequality (\ref{chi-ineq}) and lower semicontinuity of the function
$\Phi\mapsto\chi(\Phi(\mu))$ imply the second  relation in
(\ref{lim-r-2}).

Since $\{a_n\}$ obviously tends to $H(\rho)$, (\ref{I-n}),
(\ref{lim-r-1}) and (\ref{lim-r-2}) imply  (\ref{chi-dif}).
$\square$

\vspace{15pt}

I am grateful to A.S.Holevo and to the participants of his seminar "Quantum probability, statistic, information"
(the Steklov Mathematical Institute) for the
useful discussion.

\end{document}